\documentclass[prl,a4paper,nofootinbib,
twocolumn,
]{revtex4-1}
\pdfoutput=1
\usepackage{graphicx}  
\usepackage{bbold} 
\usepackage{mathtools}
\usepackage{amsmath}
\usepackage{amsfonts} 
\usepackage{amssymb}
\usepackage{amsbsy}
\usepackage{amsfonts}
\usepackage{physics}
\usepackage{bbold}
\usepackage{slashed}  
\usepackage{color}
\usepackage{hyperref} 
\usepackage{orcidlink}
\usepackage{bbding} 

\definecolor{oucrimsonred}{rgb}{0.6, 0.0, 0.0} 
\definecolor{DarkGray}{gray}{0.4}
\definecolor{forestgreen}{rgb}{0.13,0.35,0.13}
\definecolor{ocre}{HTML}{F16723}

\usepackage{bm}
\usepackage{mathrsfs}

\def\eq#1{{Eq.~(\ref{#1})}}

\def\Tr{\mbox{Tr}\,}

\colorlet{grayline}{gray!70}
\definecolor{blueline}{rgb}{0,0.27,0.55}
\definecolor{DarkGray}{gray}{0.4}
\definecolor{Gray}{gray}{0.6}
\definecolor{oucrimsonred}{rgb}{0.6, 0.0, 0.0}
\definecolor{persianblue}{rgb}{0.11, 0.22, 0.73}
\definecolor{forestgreen}{rgb}{0.13,0.35,0.13}
 \hypersetup{colorlinks, citecolor=forestgreen, linkcolor=forestgreen, urlcolor=forestgreen}
\newcommand{\be}{\begin{equation}}
\newcommand{\ee}{\end{equation}}
\newcommand{\bea}{\begin{eqnarray}}
\newcommand{\eea}{\end{eqnarray}}
\newcommand{\nn}{\nonumber}

\newcommand*\xbar[1]{%
  \hbox{\;%
    \vbox{%
      \hrule height 0.5pt 
      \kern0.5ex
      \hbox{%
        \kern-0.25em
        \ensuremath{#1}%
        \kern-0.07em
      }%
    }%
  }%
} 
\newcommand{\com}[1]{}
\newcommand{\gsim}{\lower.7ex\hbox{$\;\stackrel{\textstyle>}{\sim}\;$}}
\newcommand{\lsim}{\lower.7ex\hbox{$\;\stackrel{\textstyle<}{\sim}\;$}} 

\newcommand{\bc}{\begin{center}}
\newcommand{\ec}{\end{center}}

\newcommand{\K}{K^{*}(892)^0}

\begin{document}

\hypersetup{citecolor = forestgreen,
linktoc = section, 
linkcolor = forestgreen, 
urlcolor = forestgreen
}

\title[]{ \Large \color{oucrimsonred} \textbf{ 
  Quantum contextuality of spin-1 massive particles} }

\author{\bf M. Fabbrichesi$^{a\, \orcidlink{0000-0003-1937-3854}}$}
\author{\bf   R. Floreanini$^{a}\, \orcidlink{0000-0002-0424-2707}$}
\author{\bf E. Gabrielli$^{{b,a,c\, \orcidlink{0000-0002-0637-5124}}}$} 
\author{\bf L. Marzola$^{{c,d\, \orcidlink{0000-0003-2045-1100}}}$}
\affiliation{$^{a}$INFN, Sezione di Trieste, Via Valerio 2, I-34127 Trieste, Italy}
\affiliation{$^{b}$Physics Department, University of Trieste, Strada Costiera 11, I-34151 Trieste, Italy}
\affiliation{$^{c}$Laboratory of High-Energy and Computational Physics, NICPB, R\"avala 10,  10143 Tallinn, Estonia}
 \affiliation{$^{d}$Institute of Computer Science, University of Tartu, Narva mnt 18, 51009 Tartu, Estonia}

\begin{abstract}
Contextuality is a fundamental property of quantum mechanics. Contrary to entanglement, which can only exist in composite systems, contextuality is also present for single entities. The case of a three-level system is of particular interest because---in agreement with the Bell-Kochen-Specker theorem---it is the simplest in which quantum contextuality is necessarily present. We verify  that the polarizations of spin-1 massive particles produced at collider experiments  indeed exhibit contextuality. To this purpose we consider $W$ gauge bosons produced in top-quark decays, $J/\psi$ and $\K$ mesons created in $B$-meson decays and $\phi$ mesons resulting from $\chi^0_c$ charmonium decays, making use of the data collected and analyzed by the ATLAS, LHCb and BESIII collaborations, respectively. The polarizations of all these four particles show contextuality with a significance of more than $5\sigma$.
\end{abstract}

\maketitle

\textit{\bf Introduction.---} Quantum contextuality is the most fundamental property of a quantum system. It
refers to the impossibility of assigning a value to the result of
a measurement of a physical property independently of  the
measurement context, namely, on what other properties are
simultaneously measured with it. Hence, contextuality is the consequence of the impossibility of assigning simultaneous values to all observables of a quantum mechanical system in a way that respects their algebraic structure. The Bell-Kochen-Specker theorem~\cite{Bell:1964fg,Kochen:1968zz} shows that  a context-independent classical description of the predictions
of quantum theory is indeed impossible---at least for Hilbert spaces of dimension larger than two. Accordingly, hidden-variable models must necessarily entail the same degree of contextuality.

Quantum contextuality is a phenomenon that combines
many  aspects of quantum theory in a single
framework: from the impossibility of performing simultaneous measurements of arbitrary observables, to Bell nonlocality~\cite{Bell:1964} and entanglement~\cite{Horodecki:2009zz,Benatti:2010,Nielsen:2012yss,bruss2019quantum} when examining a composite system. Quantum contextuality has been experimentally tested at low energies by means of photons and solid-state devices~\cite{PhysRevLett.84.5457,PhysRevLett.90.250401,PhysRevLett.97.230401,PhysRevLett.103.040403,PhysRevLett.103.160405,Lapkiewicz:2011rpt}. A more detailed description of the phenomenon, and a more comprehensive list of experimental tests, can be found in two recent review articles~\cite{thompson2013,Budroni:2021rmt}.

The smallest Hilbert space in which contextuality is necessarily present is that of dimension three, describing a qutrit. Qutrits are made accessible at collider experiments by the spin degrees of freedom of massive spin-1 particles, copiously produced in scattering events. Particle physics is then the ideal setting  for an experimental test of contextuality at high energies and in the presence of weak and strong interactions.

In this Letter we analyze data about $W$ gauge bosons produced in top-quark decays, $J/\psi$ and $\K$ mesons resulting from $B$-meson decays and $\phi$ mesons created in charmonium decays, respectively collected by the ATLAS, LHCb and BESIII experimental collaborations. We find that all these four particles show contextuality with a significance of more than $5\sigma$.

\vskip0.3cm
\textit{\bf Contextuality inequalities.---}
Non-contextual hidden variable models attribute set values to physical observables, independently of the contextual 
measurement provided by additional observables. This is not possible in general if the mutual algebraic relations among these observables predicted by quantum mechanics are to be maintained. Contextuality is usually asserted by using a set of $i=1,\ldots ,N$ dichotomic observables ${\cal O}_i$, with output $\expval{{\cal O}_i}=\pm 1$, built so that the operator ${\cal O}_i$ commutes with ${\cal O}_{i+1}$, but in general does not with the remaining ones. These observables are then measured in pairs, thereby providing the different contexts necessary to rule out non-contextuality, which would necessarily predict a fixed outcome for each $\expval{{\cal O}_i}$.  The dichotomic observables of interest are easily built out of projector operators $\Pi_i = \dyad{v_i}$ as ${\cal O}_i = 1-2\,\Pi_i$, where the corresponding projections must satisfy $\langle v_i |v_{k}\rangle=0$ if $k=i+1$, in order to provide the required contexts. A graph in which the latter is rendered as a chain of vertices, each corresponding to a dichotomic operator, can be obtained by attaching to each graph vertex a ket $\ket{v_i}$, with the vectors $v_i$ of norm one in the three-dimensional real space (see the review articles~\cite{thompson2013,Budroni:2021rmt} for a more completed discussion of the procedure).

Non-contextuality is verified if the sum of the expectation values of the $N$ involved projection operators satisfies the following inequality:
\be
\boxed{\mathbb{CNTXT}_N  \equiv  \sum_{i=1}^N \langle  \Pi_i  \rangle
 \leq c_N\,,} \label{eq:def}
\ee
in which the expectation value has to be taken by means of the polarization matrix  of the particle being probed. The values  $c_5 = 2$, $c_9=3$ and $c_{13} = 4$ are found by considering  the largest number of possible insertions of projectors with the same outcome value in the corresponding contextuality chain~\cite{PhysRevLett.112.040401}. The case corresponding to a chain of 5 operators is equivalent to the 
Klyachko-Can-Binicioglu-Shumovsky (KCBS) inequality~\cite{Klyachko:2008zz}. The case of 9 projectors was introduced in Ref.~\cite{PhysRevA.86.042125}, whereas 13 operators were used by S.\ Yu and C.H.\ Oh in their proof of the Bell-Kochen-Specker theorem, which was originally based on 117 unit vectors~\cite{Kochen:1968zz}. The inequalities in \eq{eq:def} have an upper limit as well: in Quantum Mechanics it is $\sqrt{5}$ for the KCBS case, 10/3 for the case of 9 operators and 13/3 for the 13 operator one.

A qutrit described by a pure state is generically expected to violate the non-contextual inequality derived from a graph with just 5 vertices~\cite{Klyachko:2008zz}. More complicated states show contextuality using a graph of 9 vertices~\cite{PhysRevA.86.042125}, whereas a fully incoherent mixture of the three basis states will show contextuality only if a graph containing 13 vertices is used. This last case is special since the sum of the involved projection operator is proportional to the identity matrix and, therefore, the contextuality test becomes independent of the state used to compute the expectation values $\expval{{\cal O}_i}$. However, to directly assess this possibility in a laboratory, the experiment should directly probe the operator algebra---which is something impossible for the current collider detectors. 

In the following we then focus solely on the cases comprising operator chains of length 5 and 9; for the former, the set of 5 unit vectors representing the projection subspaces can be taken to be~\cite{ahrens2013fundamentalexperimentaltestsnonclassicality}
\be
|v_j\rangle  = \Big( \cos \phi, \sin \phi \, \cos \frac{4 \pi j}{5}, \sin \phi \, \sin \frac{4 \pi j}{5} \Big)^T \quad 
\ee
in which
\be
 \quad \cos^2 \phi = \frac{\cos \pi/5}{1 + \cos \pi/5}\, ,
\ee
with $j=1,2,3,4$ and 5. The set of 9 unit vectors for the complementary case is instead given by~\cite{PhysRevA.86.042125}
\begin{align}
|v_1 \rangle  & = (1,0,0)^T, \nn \\
|v_2 \rangle  & = (0,1,0)^T, \nn \\
 |v_3 \rangle  & = (0,0,1)^T,  \nn \\
|v_4 \rangle  & =(0,1,-1)^T/\sqrt{2} ,\nn\\
|v_5 \rangle  & =  (1,0,-\sqrt{2})^T/\sqrt{3}, \nn\\
 |v_6 \rangle  & = (1,\sqrt{2},0)^T/\sqrt{3}, \nn  \\
|v_7 \rangle  & = (\sqrt{2},1,1)^T/2,\nn\\
|v_8 \rangle  & = (\sqrt{2},-1,-1)^T/2, \nn\\
 |v_9 \rangle  & = (\sqrt{2},-1,1)^T/2.
\end{align}
Different choices are discussed in the review articles~\cite{thompson2013,Budroni:2021rmt}.
\com{
\begin{figure}[ht!]
\begin{center}
\includegraphics[width=3in]{./chains.png}
 \caption{\footnotesize Graphs of five and nine projector operators showing the classical exclusivity conditions. The dark vertices correspond to the possible values of 1 assigned in a context independent manner.  A gray vertex represents the  value 0.
\label{fig:chains} 
}
\end{center}
\end{figure}
}

The result of \eq{eq:def} can be optimized by finding a unitary matrix $V$ such that $[\Pi', \rho]=0$, with $\Pi' = V\Pi V^\dagger$ and $\Pi=\sum_i \Pi_i$. The trace is then to be taken after the eigenvalues of the two matrices are likewise ordered by sorting the eigenvectors in $V$~\cite{PhysRevA.86.042125}.

\vskip0.3cm
\textit{\bf Testing contextuality with spin-1 massive particles.---} 
Spin-1 massive particles are the simplest physical system for testing the presence of contextuality. The gauge bosons are the only fundamental spin-1 massive particles in the Standard Model (SM). Experimental data are available for $W$ gauge bosons, as well as for the spin-1 $J/\psi$, $\K$ and $\phi$  mesons. The  experimental analyses provide data sufficient to reconstruct the polarization  matrices of the particles by means of quantum state tomography. We use these states to estimate the expectation value of the projection operators $\Pi_i$. These matrices either directly describe the spin-1 particle, or the production in association with some other particle. In the latter case, a partial trace of the polarization matrix of the final state gives the polarization matrix of the single particle, with entries written in terms of those of the bipartite system. 

To evaluate the uncertainty affecting the expectation values in \eq{eq:def}, we numerically construct an ensemble of spin-1 states---described by properly defined polarization matrices---by sampling the matrix elements from normal distributions with standard deviations and means set to the corresponding experimental uncertainties and best fits, respectively. The final uncertainties are reduced with respect to those in the input values because the normalization of the trace of the polarization matrix strongly correlates their values.

\vskip0.3cm
\textit{\bf $W$ gauge bosons in  $t\to W^+b$  decays.---}
The $3\times 3$ polarization  matrix of the $W^+$ in the $t\to W^+ b$ decay is given in the SM at tree level by~\cite{Aguilar-Saavedra:2010ljg}
\be
{\small
\rho_W \propto 
\begin{pmatrix} 
 0 & 0 &  0 \\
  0& |m_{_{00}}|^2 (1+\cos\theta) & m_{_{0-1}} m_{_{-10}}^* \,\sin\theta \,e^{-i \varphi} \\
  0&m_{_{0-1}}^*m_{_{-1 0}} \,\sin\theta\, e^{i \varphi}& |m_{_{-1-1}}|^2 (1-\cos\theta)
  \end{pmatrix}}
\label{eq:rhoW}
\ee
on the  basis $\{+, 0 , -\}$ for the one-particle  helicity amplitudes $m_{_{n m}}$ and with a proportionality factor ensuring that $\Tr(\rho_W)=1$. The same matrix is found for the $W^{-}$ produced in the decays of the antitop quarks. The angles $\theta$ and $\varphi$ define the direction of the top quark spin with respect to the direction of its momentum. In the SM, the polarization matrix in \eq{eq:rhoW} is pure, $\Tr\qty(\rho_W^2)=1$, because $|m_{_{00}} 
| |m_{_{-1-1}}| =|m_{_{0-1}}| |m_{_{10}}|$. Hence the optimization procedure  in the estimate of contextuality gives for the SM prediction the largest eigenvalue of the projector sum in \eq{eq:def}. 

We  use \eq{eq:rhoW} to guide us in using the partial results of the experimental analysis. The polarization fractions of the $W$ gauge boson are defined as
\be
F_{i} \equiv \frac{\Gamma_{i} (t\to Wb)}{\Gamma_T(t\to Wb)}\, ,\quad \text{with}\quad  i=0,+,-
\ee
in which $\Gamma_{i}$ and $\Gamma_T$ are the corresponding polarized and total widths. They can be measured from  the decays of pairs of top quarks produced at the LHC in proton-proton collisions at a center-of-mass energy of $\sqrt{13}$ TeV. The measurement is performed selecting $t\bar t$ events decaying into final states with
two charged leptons (electrons or muons) and at least two $b$-tagged jet. The ATLAS Collaboration finds~\cite{ATLAS:2022rms}
$F_0 = 0.684 \pm 0.005|_{\text{stat}} \pm 0.014|_{\text{syst}}$, $F_- = 0.318 \pm 0.003|_{\text{stat}} \pm 0.008|_{\text{syst}}$ and  $F_+= -0.002\pm 0.002|_{\text{stat}}\pm 0.014|_{\text{syst}}$.

The coefficients $m_{_{00}}$, $m_{_{-1-1}}$ and  $m_{_{11}}$ (though the latter  vanishes in the SM, we set it  to its experimental value)  are given in terms of the helicity fractions as
\bea
|m_{_{00}}|^2 &=& F_{0} =  0.684 \pm 0.015\, ,\nn \\ 
 |m_{_{-1-1}}|^2 &=& F_-=0.318 \pm 0.009\, ,\nn\\
  |m_{_{11}}|^2&=&F_+= -0.002\pm 0.014\label{rho_exp}\, ,
\eea
with $F_0+F_-+F_+=1$. In \eq{rho_exp} we have combined statistical and systematic uncertainties in quadrature.

The lack of experimental data on the off-diagonal terms in \eq{eq:rhoW} makes it impossible to diagonalize the polarization matrix and maximize the violation of the non-contextuality bound. We then use the matrix as it is, diagonalizing only the projector sum so that the unknown off-diagonal terms do not contribute to the result of \eq{eq:def}. The values we obtain therefore are lower bounds for the non-contextuality of the involved particles.

\begin{figure}[ht!]
\begin{center}
\includegraphics[width=2.8in]{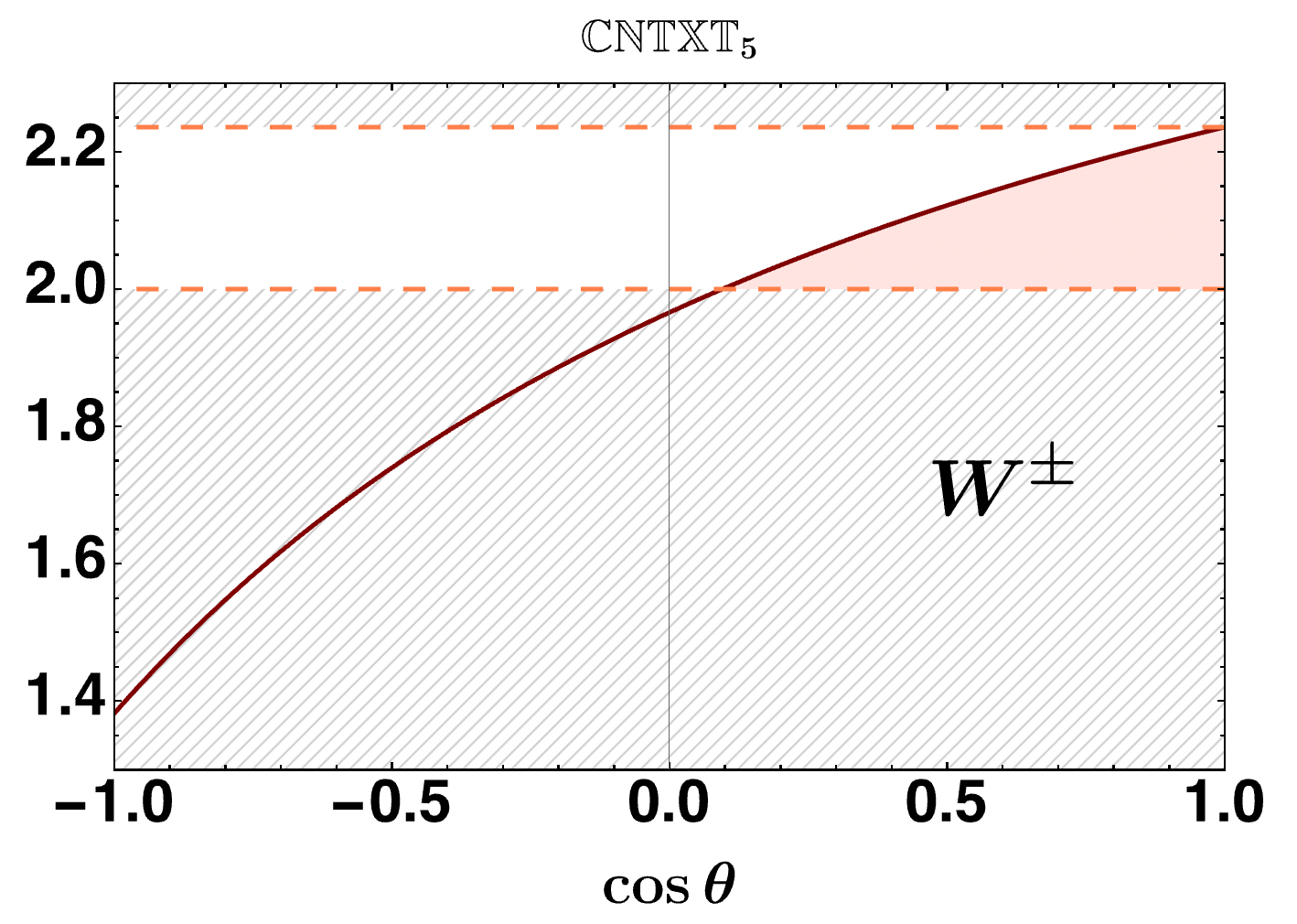}
\vskip0.3cm
\includegraphics[width=2.8in]{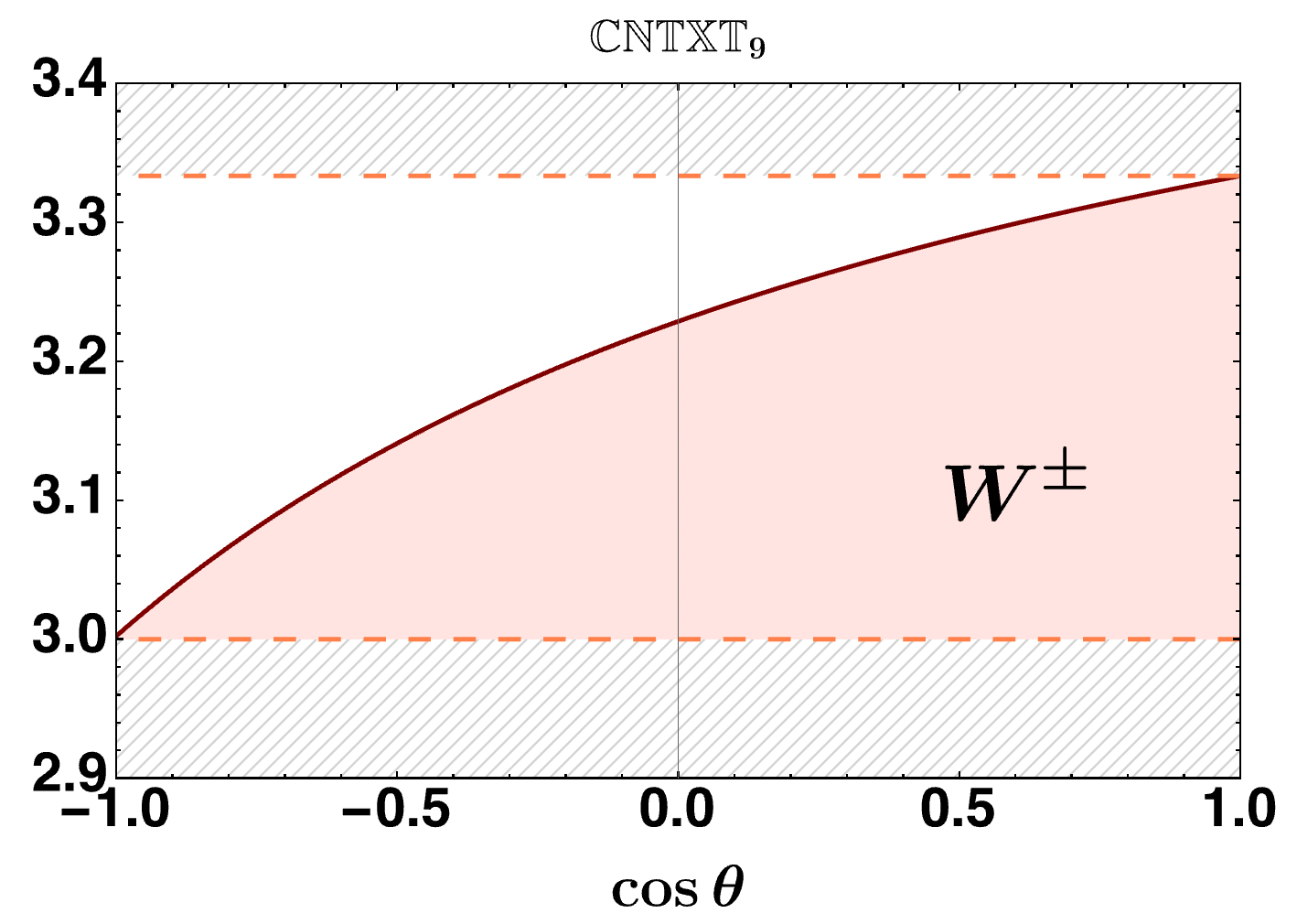}
 \caption{\footnotesize The curves plot $\mathbb{CNTXT}_5$ and  $\mathbb{CNTXT}_9$ for the $W$ gauge boson as a function of the cosine of the angle $\theta$ defined by the direction of the top-quark momentum and that of its spin. The horizontal, dashed lines mark, respectively,  the maximal value achievable assuming non-contextuality (lower line)  and the upper limit for contextuality within Quantum Mechanics (upper line).
\label{fig:context_W} 
}
\end{center}
\end{figure}

Figure~\ref{fig:context_W} shows the values of $\mathbb{CNTXT}_5$ and  $\mathbb{CNTXT}_9$ for a sample of $W$ produced in top quark decays, computed with the central values in \eq{rho_exp}, as function of the cosine of the angle $\theta$.   Whereas the non-contextuality bound is exceeded only in the forward direction for $\mathbb{CNTXT}_5$, it is for all angles in the case of  the less restrictive $\mathbb{CNTXT}_9$.

Taking the angle $\theta=\pi/4$ as a benchmark value, we find 
\bea
\mathbb{CNTXT}_5&=&2.175\pm 0.015\, , \nn \\
 \mathbb{CNTXT}_9 &=& 3.311 \pm  0.012\, ,\label{CNTXT_W}
 \eea
and a violation of the contextuality inequality with a significance well above $5\sigma$ for both of the considered operator chains.  

The contextuality test above has been performed by partially using the SM  result about the coupling between the top and the $W$ gauge boson. A test completely independent of the SM will be possible as soon as the experimental collaborations provide data to reconstruct the complete polarization matrix.

\vskip0.3cm
\textit{\bf $J/\psi$ and $\K$ mesons in $B\to J/\psi K^*$ decays.---}
The polarizations of the $J/\psi$ and $\K$ mesons directly produced in $pp$ collision are consistent with vanishing values (see, for instance, \cite{CDF:2007msx}). Polarized particles are instead produced in $B$-meson decays, hence we use the available data on the $B^0\to J/\psi \, \K$ decay~\cite{LHCb:2013vga}, collected in $p p$ collisions at 7 TeV (part of run 1 of the LHC) with the LHCb detector with an integrated luminosity of 1 fb$^{-1}$. The branching fraction for this decay is $(1.27\pm0.05)\times 10^{-3}$~\cite{ParticleDataGroup:2022pth}.

The selection of $B^{0}\to J/\psi \,\K$ events, as explained in \cite{LHCb:2013vga}, is based upon the combined decays $J/\psi \to \mu^{+}\mu^{-}$ and the $\K  \to K^{+}\pi^{-}$.  The polarizations of the spin-1 massive particles  $J/\psi$ and  $\K$ can be then reconstructed by using the momenta of the final state charged mesons and leptons produced in the secondary decays.
The analysis in Ref.~\cite{LHCb:2013vga} gives the two complex polarization amplitudes, $A_{\parallel}$ and $A_{\perp}$, as well as the corresponding phases:
\begin{align}
|A_{\parallel}|^{2} &=  0.227 \pm 0.004|_{\rm stat}\;  \pm 0.011|_{\rm sys} \nn\\
|A_{\perp}|^{2} &=  0.201 \pm 0.004|_{\rm stat}\;  \pm 0.008|_{\rm sys} \nn \\
\delta_{\parallel} \; [\text{rad}] &=  -2.94 \pm 0.02|_{\rm stat}\; \pm 0.03|_{\rm sys}  \nn\\
\delta_{\perp}  \;  [\text{rad}] &= 2.94 \pm 0.02|_{\rm stat}\; \pm 0.02|_{\rm sys}   \, ,\label{data0}
\end{align}
with $|A_{0}|^{2}+|A_{\perp}|^{2}+|A_{\parallel}|^{2}=1$, and we can take $\delta_0=0$ because there are only two physical phases. The helicity amplitudes are mapped into the polarization amplitudes used in \eq{data0} by the correspondence $w_{00}= A_{0}$, $w_{11} = (A_{\parallel}+A_{\perp})/\sqrt{2}$ and $w_{-1-1}= (A_{\parallel} -A_{\perp})/\sqrt{2}$
\com{\be
\frac{h_{0}}{|H|}= A_{0} \,, \quad \frac{h_{+}}{|H|}  = \frac{ A_{\parallel}+A_{\perp}}{\sqrt{2}}\, , \quad
\frac{h_{-}}{|H|}  = \frac{ A_{\parallel} -A_{\perp}}{\sqrt{2}} \, ,
\ee}
with $|w_{_{-1\, -1}}|^2 + |w_{_{0\,0}}|^2 + |w_{_{1\,1}} |^2 =1 $. The polarization matrix of the $J/\psi$ mesons is obtained from the polarization matrix for the whole process~\cite{Fabbrichesi:2023idl} upon a partial trace over the $K^*$ meson state as
\be
{\small
\rho_{J/\psi}= \Tr_{\!K^*}( \rho_{J/\psi K^*} )= 
  \begin{pmatrix} 
   |w_{_{1 1}}|^2 & 0 & 0  \\
  0 &|w_{_{0 0}}|^2  & 0   \\
  0 & 0 &  |w_{_{-1 -1}}|^2  \\
\end{pmatrix} }\label{rhoJpsi}
\ee
in the helicity basis of the $J/\psi$.

Adding again the statistical and systematic errors in quadrature, we find 
\bea
 \mathbb{CNTXT}_5&=&1.87\pm0.01\,, \nn \\
  \mathbb{CNTXT}_9& =& 3.18\pm 0.01\, , \label{eq:cntxt2}
\eea
with a violation of the contextuality inequality with a significance of more than $5\sigma$ with 9 operators and no violation with 5.

Since the $\K$ is also produced in the same decay, we can also find the contextuality for these mesons by simply taking the partial trace of the polarization matrix of the final state over the $J/\psi$ system. 
The polarization  matrix of the $K^*$ is just that of the $J/\psi$ with two diagonal entries exchanged, hence we have the same values for the contextuality test as in \eq{eq:cntxt2}.

Similar results can be found by analyzing data on the decays $B\to  \K \K$~\cite{LHCb:2015exo} and $B\to \phi \K$~\cite{Belle:2005lvd,LHCb:2014xzf}.

\vskip0.3cm
\textit{\bf $\phi$ mesons in $\chi^0_c \to \phi \phi$  decays.---}
The scalar   state of the charmonium $\chi^{0}_c$ can decay into a pair of spin-1 $\phi$ mesons
\be
\chi^{0}_c   \to \phi + \phi \,,
\ee
with a branching fraction of $(8.48\pm0.26 \pm 0.27)\times 10^{-4}$~\cite{BESIII:2023zcs}.
The $\chi^{0}_c$ are produced in
\be
e^{+} e^{-} \to \psi (3686) \to \gamma \,\chi^{0} \, .
\ee

The polarization  matrix of either $\phi$ meson is obtained from the polarization matrix for the whole process~\cite{Fabbrichesi:2024rec} through a partial trace as
$\rho_\phi = \Tr_{\! \phi}( \rho_{\phi\phi})$, which yields a matrix of the same form as that in \eq{rhoJpsi}, with $w_{_{1 1}} = -w_{_{-1 -1}}$ because of the conservation of parity, as well as because the final state contains indistinguishable particles. There is therefore only one independent amplitude and the polarization matrix depends on one complex number.

The analysis of the data in~\cite{BESIII:2023zcs} selects $2701\pm84$ out of the $\gamma K^+K^-K^+K^-$ final states events. The maximum likelihood fit yields the absolute value of the ratio of the moduli of the helicity amplitudes:
\be
\left| \frac{w_{_{1 1}}}{w_{_{0 0}}}\right|=0.299\pm0.003|_{\rm stat} \, \pm 0.019|_{\rm syst} \, . \label{exp1}
\ee
No value for the relative phase is provided. Accordingly, we  carry out the analysis in the case of vanishing phase, finding 
\bea
\mathbb{CNTXT}_5&=&2.11\pm0.01\,,\nn  \\
 \mathbb{CNTXT}_9 &=&3.26 \pm  0.01\, ,
\eea
with a violation of the contextuality inequality with a significance of better than $5\sigma$.

\vskip0.3cm
\textit{\bf Outlook.---}
We have established the presence of contextuality for  spin-1 massive particles by considering the analyses of the experimental data for several candidates: the $W$ gauge boson, the vector mesons $J/\psi$, $\K$ and $\phi$. These results complement those on entanglement and Bell locality for events at colliders (see, for instance, the review~\cite{Barr:2024djo} and the works cited therein) in the study of Quantum Mechanics at high energies and in the presence of strong and electroweak interactions. Quantum Mechanics is confirmed and non-contextual hidden variable models ruled out with a significance of better than $5\sigma$.  There seems to be no obvious loopholes in this  result because it is based on the experimental determination of the polarization matrices by means of quantum  state tomography.

\vskip0.3cm
\textit{Acknowledgements---}
{\small
LM is supported by the Estonian Research Council under the RVTT3, TK202 and PRG1884 grants.}
\vskip1cm
\small
\bibliographystyle{JHEP}   
\bibliography{context.bib} 

\end{document}